\documentstyle[preprint,aps,prl,epsf]{revtex}
\begin{document}
\preprint
\widetext
\title{Competition between electron-phonon attraction and weak Coulomb
repulsion}
\author{
J. K. Freericks$^{(a)}$ and Mark Jarrell$^{(b)}$.
}
\address{
$^{(a)}$Department of Physics,
Georgetown University, Washington, DC 20057-0995\\
$^{(b)}$ Department of Physics,
University of Cincinnati, Cincinnati, OH 45221
}
\date{\today}
\maketitle
\begin{abstract}
The Holstein-Hubbard model is examined in the limit of infinite dimensions.
Conventional folklore states that charge-density-wave (CDW) order
is more strongly affected by Coulomb repulsion than superconducting
order because of the pseudopotential effect.  We find that both
incommensurate CDW and superconducting phases are stabilized by the
Coulomb repulsion, but, surprisingly, the commensurate CDW transition
temperature is more robust than the superconducting transition
temperature.  This puzzling feature is resolved by a detailed analysis
of perturbation theory.
\end{abstract}
\renewcommand{\thefootnote}{\copyright}
\footnotetext{ 1994 by the authors.  Reproduction of this article by any means
is permitted for non-commercial purposes.}
\renewcommand{\thefootnote}{\alpha{footnote}}
\pacs{74.20.-z, 71.27.+a, 71.38.+i}

\narrowtext
\paragraph*{Introduction.}
Interacting electronic systems with both electron-electron repulsion and
phonon-mediated attractive interactions display a rich variety of ground
states due to the competition between these interactions.  In conventional
low-temperature superconductors (SC), the electron-phonon interaction dominates
over the Coulomb repulsion, since the latter
is reduced from its bare values by the so-called pseudopotential effect (as
described by Bogoliubov, et. al. \cite{bogoliubov}
and Morel and Anderson \cite{morel}).  The electron-phonon interaction is also
believed to be responsible for charge-density-wave (CDW) order in weakly doped
BaBiO$_3$.  However, the Coulomb repulsion dominates over the electron-phonon
interaction in the high-temperature superconducting cuprates, with commensurate
spin-density-wave (SDW) order occurring in the (undoped) parent compounds.

Although the Coulomb interaction (in the form of the Hubbard model
\cite{jarrell,kotliar}) and the electron-phonon interaction (in the form of
the Holstein model \cite{freericks_qmc}) have both been extensively studied,
there has been only limited work on the combined Holstein-Hubbard model
\cite{schuttler} and only one exact theorem for the case of an attractive
Coulomb interaction \cite{freericks_lieb}.

The conventional lore for the effect of the Coulomb interaction on a
superconductor with strong electron-phonon interactions, is that the
Coulomb repulsion is {\it reduced} from its bare values.  This is because the
electron-phonon interaction is {\it retarded}, allowing the electrons to
attract each other, through the exchange of a virtual phonon, without being
at the same lattice site at the same time.  A quantitative estimate for this
so-called pseudopotential effect can be made, and the (dimensionless) Coulomb
repulsion $\rho(\mu)U_c$ is reduced to
\begin{equation}
\rho(\mu)U_c^*:=\frac{\rho(\mu)U_c}{1+\rho(\mu)U_c\ln {W\over 2\omega_D}}\quad
,
\label{eq: pseudo_lore}
\end{equation}
where $\rho(\mu)$ is the electronic density of states (DOS) for an electron
(of one spin) at the Fermi energy, $U_c$ is the bare Coulomb repulsion, $W$
is the electronic bandwidth, and $\omega_D$ is the Debye frequency.  There is,
however, no pseudopotential effect for a CDW distortion, because retardation
effects play a limited role in a {\it static} CDW, where the electrons
remained paired at every other lattice site.
Therefore, we expect that the Coulomb repulsion will reduce the transition
temperatures for CDW order much more than for SC, and that the SC phase is
thereby stabilized relative to the CDW phase.

\paragraph*{Formalism.} The dynamical mean-field theory (MFT) of Metzner
and Vollhardt \cite{metzner_vollhardt} has been employed to exactly solve
the Hubbard \cite{hubbard} and Holstein \cite{holstein} models in infinite
spatial dimensions using the quantum Monte Carlo (QMC) algorithm of
Hirsch and Fye \cite{hirsch_fye}.  The Holstein-Hubbard model
\begin{eqnarray}
H&=&-{t^*\over 2\sqrt{d}}\sum_{<i,j>\sigma}c_{i\sigma}^{\dagger}c_{j\sigma}
+g\sum_ix_i(n_{i\uparrow}+n_{i\downarrow}-1)+U_c\sum_in_{i\uparrow}
n_{i\downarrow}\cr
&+&{1\over 2}\sum_i ({p_i^2\over M}+M\Omega^2x_i^2)\quad ,
\label{eq: hdef}
\end{eqnarray}
is written in standard notation: $c_{i\sigma}^{\dagger}$ is a creation
operator for an electron localized at site $i$ with spin $\sigma$;
$n_{i\sigma}:=c_{i\sigma}^{\dagger}c_{i\sigma}$ is the corresponding number
operator; $x_i$ is the phonon coordinate at site $i$; and $p_i$ is its
momentum.  The hopping integral is $t=:t^*/2\sqrt{d}$, the deformation
potential
is $g$, $U_c$ is the Coulomb repulsion, $\Omega$ is the phonon frequency, and
$M=1$ is the phonon mass.  The scaled hopping integral $t^*$ determines the
energy unit and is set equal to one $(t^*=1)$.  The effective electron-electron
attraction (due to phonon exchange) satisfies $U=-g^2/M\Omega^2$, and competes
with the Coulomb repulsion $U_c>0$.

The Holstein-Hubbard model is solved by QMC simulation.
The QMC algorithm determines the local electronic Green's function
$G(i\omega_n)$ at each Fermionic Matsubara frequency $\omega_n:=\pi T(2n+1)$,
by mapping the infinite-dimensional lattice to an impurity problem.  This
mapping has been described in detail elsewhere \cite{jarrell,infd_imp}.
The Green's function satisfies
\begin{equation}
G(i\omega_n)=F_{\infty}[i\omega_n+\mu-\Sigma(i\omega_n)]\quad ,
\label{eq: gdef}
\end{equation}
with $\Sigma(i\omega_n)$ the electronic self energy and $F_{\infty}(z):=
\int dy \rho(y)/(z-y)$ the rescaled complementary error function of a
complex argument [$\rho(y):=\exp (-y^2)/\sqrt{\pi}$ is the noninteracting
DOS].

The momentum-dependent susceptibility $\chi({\bf q})=$ $T\sum\tilde\chi_{mn}(
{\bf q})
=:$ $ T\sum\tilde\chi({\bf q},i\omega_m,i\omega_n)$ (for CDW, SDW, or SC order)
satisfies a Dyson equation
\begin{equation}
\tilde\chi_{mn}({\bf q})=\chi^0_m({\bf q}
)\delta_{mn}-T\sum_p\chi^0_m({\bf q})\Gamma_{mp} \tilde\chi_{pn}({\bf q})
\label{eq: chi_dyson}
\end{equation}
for each ordering vector {\bf q}, with $\chi^0$ the relevant bare
susceptibility,
and $\Gamma_{mp}$ the local irreducible vertex function. The bare
susceptibility
for commensurate [${\bf q}={\bf Q}:=(\pi,\pi,\pi,...)$] CDW or SDW order is
\begin{equation}
\chi^0_n({\bf Q}):=-\frac{G(i\omega_n)}{i\omega_n+\mu-\Sigma(i\omega_n)}\quad ,
\label{eq: chi0cdwdef}
\end{equation}
with a more complicated form for
incommensurate wavevectors \cite{mueller_hartmann}.
On the other hand, the uniform bare susceptibility for SC order satisfies
\begin{equation}
\chi^{0\prime}_n({\bf 0}):=-\frac{{\rm Im}G(i\omega_n)}{\omega_n-{\rm Im}
\Sigma(i\omega_n)}\quad .
\label{eq: chi0scdef}
\end{equation}
The irreducible vertex functions are extracted directly from the QMC data
\cite{freericks_qmc}.  Figure~1 displays the lowest-order diagrammatic
contributions to the vertex functions in the (a) CDW, (b) SC, (c) and SDW
channels.

Transition temperatures are found by starting in the disordered
(high-temperature) phase, and reducing the temperature until the susceptibility
for each ordered phase diverges.  The highest transition temperature $T_c$
determines the initial symmetry of the ordered phase.

\paragraph*{Results.} The phase diagrams for the Holstein-Hubbard model
with $g=0.5$, $\Omega=0.5$, $U=-1.0$, and $U_c=0.0$, 0.25, 0.5, 0.75, are
displayed in Figure~2.  These phase diagrams are determined by QMC calculations
and by a second-order iterated perturbation theory (IPT) \cite{freericks_ipt}
(there are no detectable phase transitions with the IPT for $U_c=0.75$).
The solid dots (lines) depict the commensurate CDW, the open dots (dotted
lines) depict the incommensurate order, and the open triangles (dot-dashed
lines) depict the SC phase for the QMC (IPT) calculations (the dashed lines
connecting the QMC points are a guide to the eye).

        The QMC data displays two types of notable behavior.  First, there
are no stable incommensurate phases when $U_c=0$; the incommensurate phases
become stable near the CDW-SC phase boundary only for $U_c>0$.  The explanation
for this is simple: if the SC phase was ignored, the CDW phase would suffer a
commensurate-incommensurate phase transition as $T_c\rightarrow 0$ in a similar
fashion to the repulsive Hubbard model \cite{freericks_maghubb};  however, the
SC transition temperature is greater than the highest incommensurate CDW
transition temperature, precluding its appearance at $U_c=0$.  As $U_c$
increases, the SC $T_c$ drops below the maximal incommensurate CDW $T_c$, which
allows the incommensurate order to occur.  Second, even though the region where
the CDW phase is stable shrinks as $U_c$ is increased, the CDW transition
temperature {\it is reduced by a smaller factor} than the SC transition
temperature, in opposition to the conventional folklore.

The IPT approximation is reasonable for both the CDW and SC $T_c$'s and for
the phase boundary between CDW and SC order. The approximate CDW transition
temperature is again
reduced by a smaller factor than the SC transition temperature
as $U_c$ increases.  Thus one can understand this effect
by studying the weak-coupling formalism.  The IPT errs only by
predicting a large incommensurate CDW-ordered region at $U_c=0$ which shrinks
as $U_c$ increases, exactly opposite to what the QMC found, and it is unable
to predict any finite $T_c$'s for $U_c=0.75$.

The modification of the CDW transition temperature at half filling
is plotted versus $U_c$ in Figure~3.  The CDW phase is followed
for $0<U_c<|U|$, since a SDW phase is expected to be the stable phase for
$U_c>|U|$ at half filling \cite{schuttler} (this can be seen at weak coupling
by comparing the CDW vertex to the SDW vertex in a power series as shown
in Figure~1, indicating $\Gamma_{SDW}>\Gamma_{CDW}$ when $U_c > |U|$).
$T_c(U_c)$ is smaller than $T_c(U_c=0)$ in the weak-coupling regime $(g<0.625)$
and the curves are nearly linear in $U_c/|U|$, with a decreasing slope as $g$
increases.  In the strong-coupling regime $(g=1.0)$, the Coulomb repulsion
initially {\it enhances} the transition temperature (since it reduces the
energy
of the virtual state formed by breaking the bipolaron) before causing a
reduction as $U_c\rightarrow|U|$.

\paragraph*{Theory.}  Much of the unexpected and notable behavior found in
the QMC and ITP results can be illustrated within an analytic approximation.
A weak-coupling analysis of the CDW and SC transition temperatures is performed
in the square-well approximation [where the soft cutoffs
$\Omega^2/\{\Omega^2+(\omega_m-\omega_n)^2\}$ are replaced by hard cutoffs
$\theta(\omega_c-|\omega_m|)\theta(\omega_c-|\omega_n|)$]
\cite{freericks_weak}.  A first-order calculation is accurate only to the
lowest
order in $1/U$.  We summarize the main results here.  Details will
be given elsewhere.

In the SC channel, for small $U_c$ one finds \cite{marsiglio}
\begin{equation}
\frac{T^{SC}_c(U_c)}{T^{SC}_c(0)}\approx \exp\left [ -\frac{1}{\rho(\mu)|U|
-\frac{\rho(\mu)U_c}{1+\rho(\mu)U_cI}}+\frac{1}{\rho(\mu)|U|}\right ]
\approx\exp\left [ -\frac{U_c/|U|}{\rho(\mu)|U|}\right ] \quad ,
\label{eq: tc_sc}
\end{equation}
for arbitrary filling, with
\begin{eqnarray}
I&:=&-\frac{T}{\rho(\mu)}\sum_{|\omega_n|>\omega_c} \frac{{\rm Im} F_{\infty}
(i\omega_n+\mu)}{\omega_n}\cr &\approx&
\frac{2}{\pi}\int_0^{\infty}\frac{dy}{y}\frac{\rho(y+\mu)+\rho(y-\mu)}{2
\rho(\mu)}\tan^{-1}\frac{y}{\omega_c}\quad ,
\label{eq: idef}
\end{eqnarray}
and $\omega_c$ is the cutoff frequency for the square well (the second line
holds when $T_c\ll \omega_c$).
In the limit $\omega_c\rightarrow 0$, one finds $I\rightarrow\ln (W/2
\omega_c)$, as in the original work \cite{bogoliubov}, but the above expression
also holds for arbitrary electronic DOS \cite{imax}.

Calculations in the CDW phase are more difficult.  Restricting to the case of
half filling $(\mu = 0)$ and again for small $U_c$, one finds
\begin{eqnarray}
\frac{T^{CDW}_c(U_c)}{T^{CDW}_c(0)}&\approx& \exp\left [ -\frac{1}
{\rho(0)[\{2-\alpha(U_c)\}|U|-U_c]}+\frac{1}{\rho(0)\{2-\alpha(0)\}|U|}\right ]
\cr
&\approx& \exp \left [ -\frac{U_c/|U|+\alpha(U_c)-\alpha(0)}
{\{2-\alpha(0)\}^2\rho(0)|U|}\right ]\quad,
\label{eq: tc_cdw}
\end{eqnarray}
with $\alpha(U_c)$ a parameter that measures the reduction of the direct
electron-phonon attraction by the exchange diagrams in Fig.~1(a).  This
parameter satisfies $0<\alpha <1$ with $\alpha\rightarrow 0$ as $\Omega
\rightarrow 0$ and $\alpha\rightarrow 1$ as $\Omega\rightarrow\infty$.
An estimate for $\alpha$ yields
\begin{equation}
\alpha(U_c)\approx 1 - \frac{I\rho(0)(|U|-U_c)}{1-I\rho(0)|U|}
\label{eq: alpha_approx}
\end{equation}
which does approach 1 as $\omega_c\rightarrow\infty$ $(I\rightarrow 0)$ and
0 as $\omega_c\rightarrow 0$ $[I\rightarrow 1/\rho(0)(2|U|-U_c)]$\cite{imax}.
Substituting Eq.~(\ref{eq: alpha_approx}) into Eq.~(\ref{eq: tc_cdw})
finally yields
\begin{equation}
\frac{T^{CDW}_c(U_c)}{T^{CDW}_c(0)}\approx \exp\left [ -\frac{U_c}{|U|}
\Big ( \frac{1}{\rho(0)|U|}-I\Big ) \right ]\quad .
\label{eq: tc_cdw2}
\end{equation}

The resolution of the puzzle of how $U_c$ affects $T_c(CDW)$ versus $T_c(SC)$
is seen by examining the small $U_c$ limits of Eqs.~(\ref{eq: tc_sc})
and (\ref{eq: tc_cdw}).  The pseudopotential effect disappears in the SC
channel as $U_c\rightarrow 0$, and the effect on $T_c$ is enhanced away from
hal
f
filling since $\rho(\mu)<\rho(0)$.  On the other hand, the effect of $U_c$ on
the CDW transition temperature is further reduced by the parameter $I$ in
Eq.~(\ref{eq: tc_cdw2}).
Thus, at least in the weak-coupling limit, $U_c$
initially reduces the SC $T_c$ more than it does the CDW $T_c$.

Fig.~3 shows the
weak-coupling results (with the parameter $I$ fitted to the QMC data for small
Coulomb repulsion).  One can see the weak-coupling formalism is excellent
for $g=0.4$, but becomes less accurate as the coupling strength increases.

In the strong-coupling limit, where the electrons are paired into bipolarons
at a temperature much higher than the transition temperature for the ordered
phase, the initial effect of $U_c$ is to {\it enhance} the CDW $T_c$, because
Coulomb repulsion {\it reduces} the bipolaron binding energy $E_b$, thereby
increasing $T_c\propto 1/E_b$.  The analysis for the pure electron-phonon case,
is easily modified by changing the energy of all intermediate states to take
into account the Coulomb repulsion \cite{freericks_strong}.  The CDW transition
temperature at half filling then satisfies
\begin{eqnarray}
\frac{T^{CDW}_c(U_c)}{T^{CDW}_c(0)}\approx \frac{1}{1-\frac{U_c}{|U|}}&\,&
\left [ 1+\sum_{n=1}^{\infty}\frac{(-S)^n}{(1+S^{\prime})(2+S^{\prime})...
(n+S^{\prime})}\right ]\cr \Big /
&\,&\left [ 1+\sum_{n=1}^{\infty}\frac{(-S)^n}{(1+S)(2+S)...
(n+S)}\right ]
\label{eq: tc_strong}
\end{eqnarray}
to second order in $|U|$, with $S:=|U|/\Omega$ and
$S^{\prime}:=S-(U_c/\Omega)$.
This analysis has been extended to fourth order, but it is too cumbersome to be
shown here.  The solid line in Fig.~3 depicts the fourth-order approximation
for
$T_c(U_c)/T_c(0)$ [which lies with 10\% of Eq.~(\ref{eq: tc_strong}) for
$U_c<0.5|U|$] when $g=1$.  Note how poor this approximation is as $U_c$ becomes
a sizeable fraction of $|U|$.

\paragraph*{Conclusions.}
We have found that the conventional folklore for how Coulomb repulsion affects
the electron-phonon interaction is flawed in assuming the SC transition
temperature is more robust than the CDW transition temperature, but is
resolved by detailed analysis of the weak-coupling theory.  The Coulomb
repulsion stabilizes the SC phase relative to the CDW phase and also allows
for the appearance of incommensurate CDW order because $T_c^{max}(SC)<T_c^{max}
(ICDW)$ for finite $U_c$.  What happens to the phase diagram for $U_c>|U|$?
We conjecture that the CDW phase is taken over by the SDW phase at half
filling,
but do not know whether or not SC phases can remain stable away from half
filling (due to the pseudopotential effect) or if paramagnetism prevails.
Work along these lines is in progress.

We would like to acknowledge useful discussions with
J.\ Deisz,
R.\ Scalettar,
D.\ Scalapino,
and H.-B. Schuttler.
This work was
supported by the National Science Foundation grant No. DMR-9107563.
In addition MJ would like to acknowledge the support of the
NSF NYI program.  Computer support was provided by the Ohio
Supercomputer Center.

\begin{figure}[t]
\epsfxsize=6.0in
\epsffile{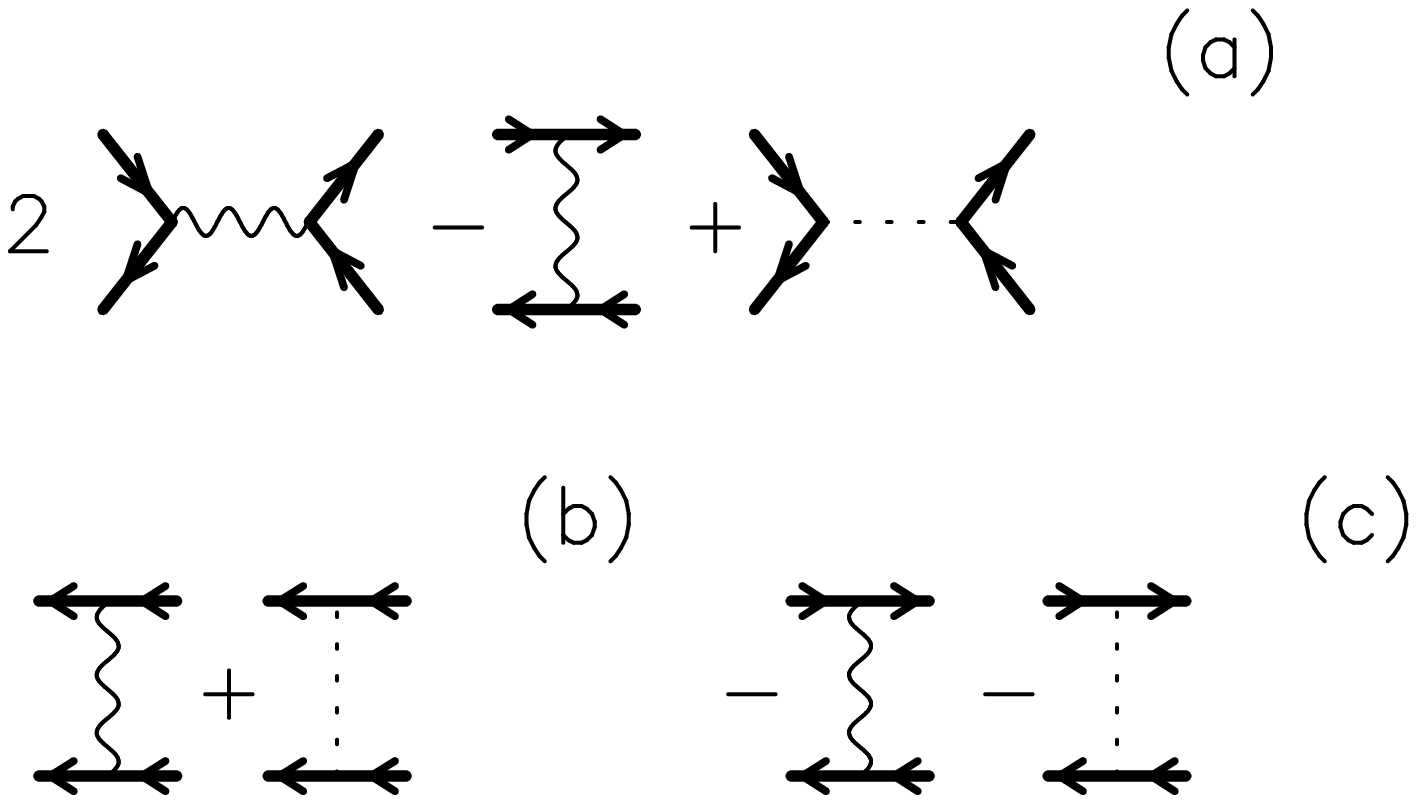}
\caption{Lowest-order contribution to the irreducible vertex function in
the (a) charge-density-wave, (b) superconducting, and (c) spin-density-wave
channels.  The solid lines denote electron propagators, the wiggly lines
denote phonon propagators, and the dotted lines denote the Coulomb repulsion. }
\end{figure}

\pagebreak

\begin{figure}[t]
\epsfxsize=6.0in
\epsffile{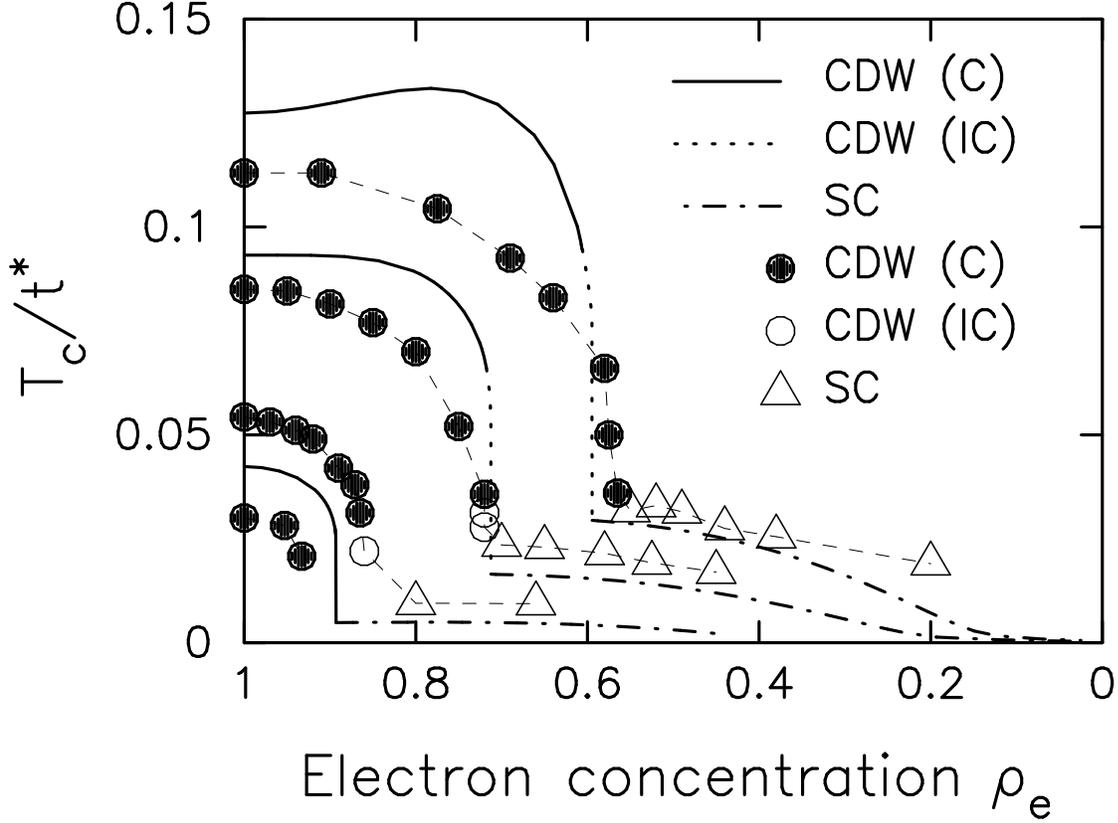}
\caption{Phase diagram for the Holstein-Hubbard model with $g=0.5$,
$\Omega=0.5$, $U=-1.0$, and $U_c=$ 0.0, 0.25, 0.5, and 0.75.  The solid dots
(lines) are for the commensurate CDW, the open dots (dotted lines) are for
the incommensurate CDW, and the open triangles (dot-dashed lines) are for the
SC as
determined from a QMC (IPT) calculation.  There are no transitions detected
by the IPT for $U_c=0.75$.  The dashed line through the QMC points is a guide
to the eye. }
\end{figure}

\pagebreak

\begin{figure}[t]
\epsfxsize=6.0in
\epsffile{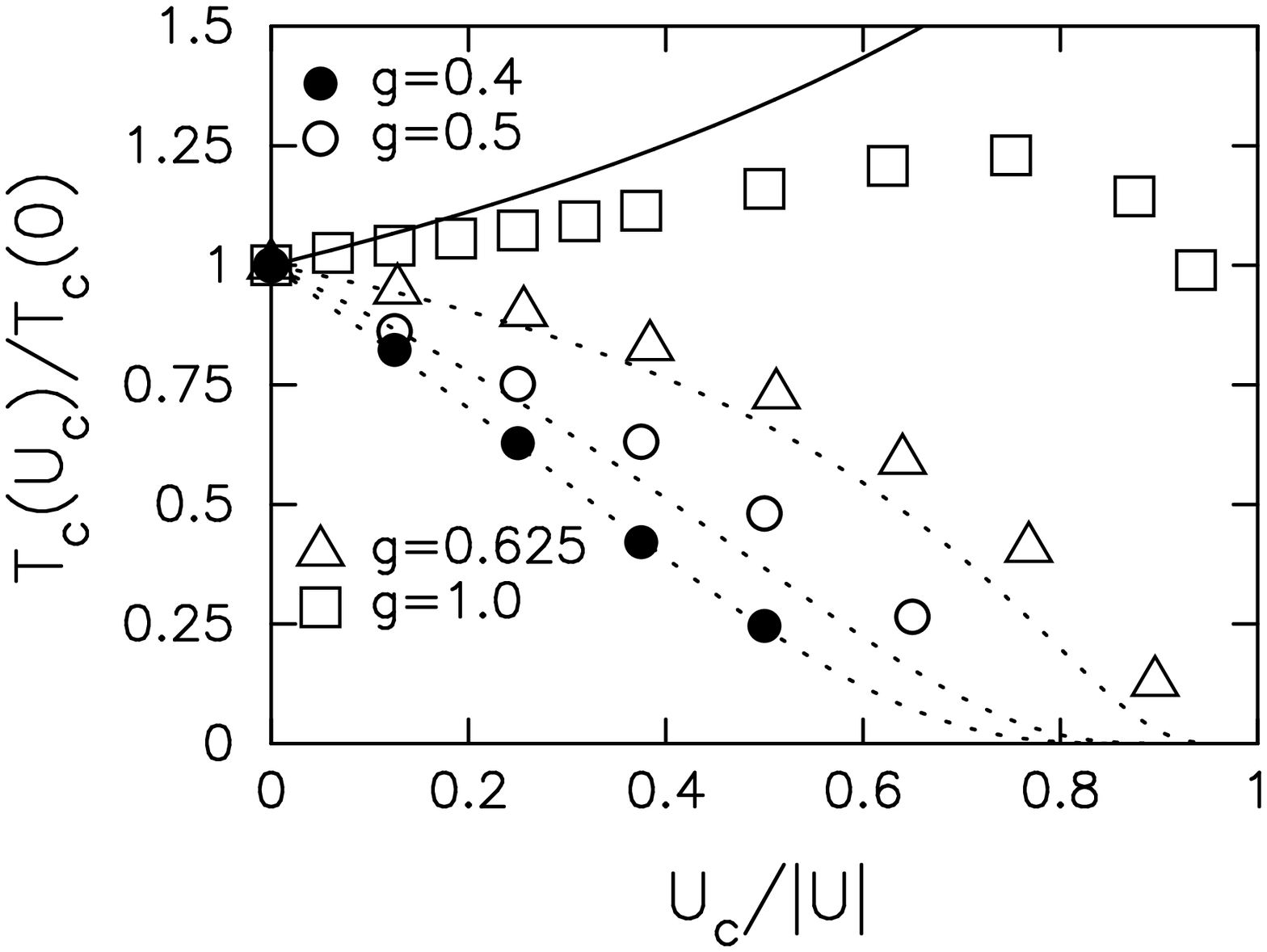}
\caption{Relative change in the CDW transition temperature at half filling
due to Coulomb repulsion. Four different cases are shown: $g=0.4$ (solid dots);
$g=0.5$ (open dots); $g=0.625$ (open triangles); and $g=1.0$ (open squares).
The dotted lines are the weak-coupling approximations for $g=0.4$, 0.5, and
0.625.  The parameter $I$ is adjusted to produce the correct slope as
$U_c\rightarrow 0$.  The solid line is a fourth-order strong-coupling
approximation for the $g=1.0$ case.}
\end{figure}


\begin{thebibliography}{10}
\bibitem{imax} In the weak-coupling limit, one can show that the maximal
value for the parameter $I$ is set by the interaction strength in the
following way: $I_{max}<2/\rho(\mu)|U|$ for the SC channel; and
$I_{max}=1/\rho(0)[(2-\alpha)|U|-U_c]$ for the CDW channel at half filling.
\end{thebibliography}
\end{document}